\documentclass[twocolumn,tighten]{aastex7}
\usepackage{amssymb, amsmath,framed}

\usepackage{bm}
\expandafter\ifx\csname package@font\endcsname\relax\else
 \expandafter\expandafter
 \expandafter\usepackage
 \expandafter\expandafter
 \expandafter{\csname package@font\endcsname}
\fi
\def\bq{\begin{equation}}
\def\eq{\end{equation}}
\def\bqy{\begin{eqnarray}}
\def\eqy{\end{eqnarray}}

\def\p{\partial}

\def\p{\partial}

\def\R{\mathbb{R}}

 \def\q#1#2{q^{\hspace{1pt}#1}_{\,,\hspace{1pt}#2}}
 \def\a#1#2{a^{\hspace{1pt}#1}_{\,,\hspace{1pt}#2}}
 \def\d#1#2{\delta^{\hspace{1pt}#1}_{\,,\hspace{1pt}#2}}

\begin{document}

\title{Surrogate Modeling of Landau Damping with Deep Operator Networks}

\correspondingauthor{Simin Shekarpaz, Chuanfei Dong}
\email{siminsh@bu.edu, dcfy@bu.edu}

\author[orcid=0009-0000-0502-5264,gname=Simin, sname='Shekarpaz']{Simin Shekarpaz}
\affiliation{Center for Space Physics and Department of Astronomy, Boston University, Boston, MA 02215, USA}
\email{siminsh@bu.edu}

\author[orcid=0000-0002-8990-094X,gname=Chuanfei, sname='Dong']{Chuanfei Dong} 
\affiliation{Center for Space Physics and Department of Astronomy, Boston University, Boston, MA 02215, USA}
\email{dcfy@bu.edu}

\author[orcid=0000-0002-8624-1264,gname=Ziyu, sname='Huang']{Ziyu Huang} 
\affiliation{Center for Space Physics and Department of Astronomy, Boston University, Boston, MA 02215, USA}
\email{zyuhuang@bu.edu}


\begin{abstract}

Kinetic simulations excel at capturing microscale plasma physics phenomena with high accuracy, but their computational demands make them impractical for modeling large-scale space and astrophysical systems. In this context, we build a surrogate model, using Deep Operator Networks (DeepONets), based upon the Vlasov-Poisson simulation data to model the dynamical evolution of plasmas, focusing on the Landau damping process - a fundamental kinetic phenomenon in space and astrophysical plasmas. The trained DeepONets are able to capture the evolution of electric field energy in both linear and nonlinear regimes under various conditions. Extensive validation highlights DeepONets' robust performance in reproducing complex plasma behaviors with high accuracy, paving the way for large-scale modeling of space and astrophysical plasmas.

\end{abstract}


\section{\textbf{Introduction}}
\label{sec1}

In space and astrophysical plasmas, most physical processes are collisionless. Generally, simulating these collisionless processes demands kinetic approaches, such as the particle-in-cell (PIC) or Vlasov methods. However, these kinetic simulations are computationally demanding and therefore not well suited for efficiently addressing large-scale problems that involve collisionless physics. To manage this challenge with affordable computational costs, two main approaches have been developed for large-scale global simulations: the magnetohydrodynamics with embedded particle-in-cell (MHD-EPIC) model \citep{Daldorff2014,Toth2016,Chen2017,Zhou2020} and the multi-moment multi-fluid model \citep{Wang2015,Wang2018,Dong2019,Wang2020,Jarmak2020,Rulke2021}. Recently, the MHD-EPIC model has evolved to include adaptively embedded PIC regions (MHD-AEPIC) \citep{Wang2022,Chen2023,Li2023}, enabling flexibility to capture localized regions where kinetic effects are significant. On the other hand, substantial efforts have been dedicated to integrating kinetic effects into the multi-moment multi-fluid model through machine learning techniques \citep{Ma2020machine,Alves2022,Cheng2023,Qin2023,Donaghy2023,Huang2025,Ingelsten2025}. 

The rapid evolution of neural network architectures has positioned machine learning as a promising approach for scientific discovery in partial differential equations (PDEs) and the development of surrogate models \citep{Ma2020machine,Laperre2022,Parand2022,Hajimohammadi2023,Razzaghi2023,Cheng2023,Joglekar2023,Qin2023,Wei2023,Shukla2025}. Early efforts, such as those by \citet{Ma2020machine}, employed surrogate models for Hammett-Perkins closure \citep{Hammett1990fluid} using various architectures, including multilayer perceptrons (MLP), convolutional neural networks (CNN), and discrete Fourier transform (DFT) networks. \citet{Raissi2019} introduced Physics-Informed Neural Networks (PINNs), which combine neural network approximation with physics-based constraints to learn solutions to differential equations directly from data. PINNs have demonstrated success in solving of ordinary and partial differential equations, including fractional equations \citep{Pang2019}, stochastic partial differential equations \citep{Zhang2020-2}, systems of differential equations \citep{Shekarpaz2024}, and inverse problems \citep{Meng2020}, without requiring explicit discretization. To further improve the robustness and accuracy of PINNs in solving high-dimensional nonlinear problems, \cite{Shekarpaz2022} proposed a physics-informed adversarial training (PIAT) framework. Adversarial training has been shown to be effective in achieving robustness against the specific perturbations used during training (\cite{Azizmalayeri2023}).
It is worth mentioning that \citet{Qin2023} utilized PINNs to construct a multi-moment fluid model with an implicit fluid closure and applied it to study the Landau damping process focusing on the linear damping case. 

Recently, operator learning - a deep learning framework that approximates linear and nonlinear differential operators by taking parametric functions (infinite-dimensional objects) as input and mapping them to complete solution fields -  becomes a hot topic, among which, Fourier Neural Operator (FNO) \citep{Li2020fourier} and Deep Operator Networks (DeepONets) \citep{Lulu2021} are probably the two most popular ones.  In this study, we will use the latter. While traditional numerical methods often rely on discretization and operator splitting \citep[e.g.,][]{Parand2019, Shekarpaz2020}, DeepONets utilize neural networks to directly approximate solutions of high-dimensional differential equations. This eliminates the need for explicit discretization or operator splitting, providing a scalable and efficient alternative to conventional numerical approaches. Notably, DeepONets enable accurate modeling of complex physical systems with strong generalization capabilities, which is an advantage over both traditional numerical solvers and other neural network-based approaches such as PINNs and standard artificial neural networks (ANNs). DeepONets have demonstrated promising performance across a wide range of applications, including weather forecasting \citep{Pathak2023}, multi-physics and multi-scale modeling \citep{Cai2021}, disk-planet interactions in protoplanetary systems \citep{Mao2023}, and optimization \citep{Sahin2024}.

In this paper, we aim to build a surrogate model using DeepONets based upon the Vlasov-Poisson simulation data to model the dynamical evolution of plasmas, focusing on the Landau damping process - a fundamental kinetic phenomenon in space and astrophysical plasmas. The trained DeepONets are able to capture the evolution of electric field energy in both linear and nonlinear regimes under various conditions. It is noteworthy that \citet{Qin2023} focused exclusively on linear Landau damping using PINNs, as the damping trend in this regime is monotonic and no electron phase-space holes are observed (see their Figure 3).

This paper is structured as follows: Section 2 introduces DeepONets. In Section 3, we describe the physical model and the dataset generation. Section 4 focuses on demonstrating the accuracy and robustness of DeepONets by applying them for the Landau damping problem under two different scenarios. Section 5 gives the conclusion.

\section{\textbf{methodology}}
\label{sec2}
\textbf{DeepONets:} 
The foundation of DeepONets is based on the universal approximation theorem \citep{Hornik1989}, which guarantees that neural networks can approximate any continuous function with arbitrary accuracy. DeepONets are designed to learn mappings between function spaces, making them useful in the context of partial differential equations (PDEs). 

\begin{figure}[htbp]
    \centering
    \includegraphics[width=0.5\textwidth]{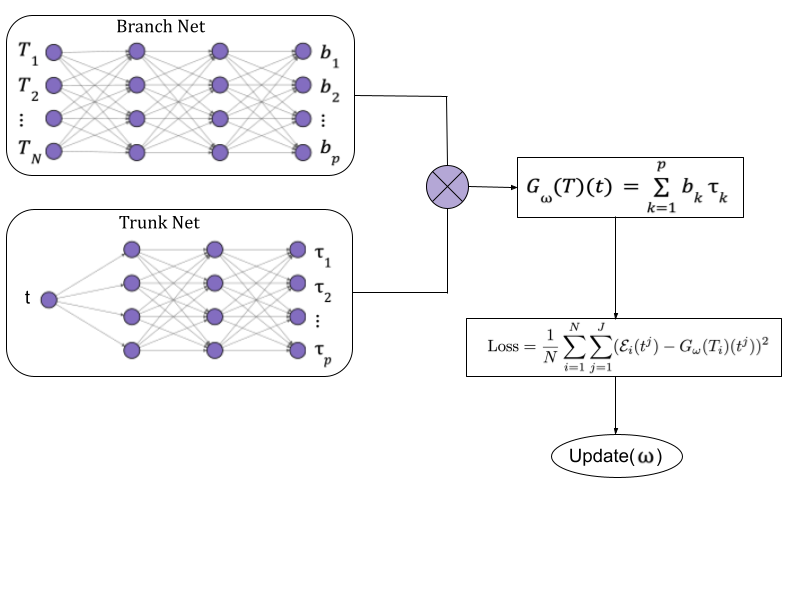}
    \caption{The DeepONets architecture consists of two fully connected neural networks (FNN): the branch network, which encodes temperature values, $T$, and the trunk network, which encodes the coordinates (here the coordinates are time, $t$). The outputs of these networks are combined to approximate the electric field energy, $\int |E_x|^2 dx$. The final output is $G_{\omega}(T)(t)$, where $\omega$ represents the model's learned weights and $T$ is a vector as $\{T_i\}_{i=1}^{N}$. }
    \label{fig:deeponet_architecture:}
\end{figure}

As a reference, consider the parametric PDEs of the form 
\begin{equation}\label{pde}
\mathcal{N}(T, \mathcal E) = 0,
\end{equation}
where $\mathcal{N}$ can be a linear or nonlinear differential operator, $T$ denotes the parametric input, and $\mathcal E$ is the corresponding functional output. DeepONets can be used to capture the relationship between $T$ and $\mathcal E$, represented as
$$\mathcal E = G_{\omega}(T)(t) \approx \sum_{k=1}^{p} b_k(T(\eta_1), T(\eta_2), \ldots, T(\eta_m)) \ \tau_k(t)$$ 
where $t$ denotes the collocation points, $\omega$ represents the network parameters, and $b_k$ and $\tau_k$ are the outputs of branch and trunk networks, as shown in Figure \ref{fig:deeponet_architecture:}. The function $T$ evaluated at fixed sensors $\{\eta_i\}_{i=1}^{m}$ will be used as the input of the branch network. For the Landau damping problem, $G_{\omega}$ maps a set of temperature values, $T$, to the electric field energy, $G_{\omega}(T)(t)$, at different time steps, $t$. 

For different kinds of problems, the branch and trunk networks can be residual network (ResNet), convolutional neural network (CNN), recurrent neural network (RNN), or feed forward neural network (FFNN).
In high-dimensional problems, where $t$ is a vector with $d$ components, the dimension of
$t$ no longer matches the dimension of $T(\eta_i)$ for $i=1, 2, \ldots, m$, and at least two sub-networks are necessary to handle 
$[T(\eta_1), T(\eta_2), \ldots, T(\eta_m)]^T$ and $t$ separately. Furthermore, depending on the number and characteristics of input functions, one can incorporate multiple branch networks instead of one \citep{Lulu2021}.\\

\section{\textbf{Implementation}}
\label{sec3}
\subsection{\label{sec:setup} Physical Model for Data Generation}
For collisionless electrostatic plasmas, their kinetic behavior can be well described by the Vlasov-Poisson equations. The Vlasov equation describes the evolution of the plasma distribution function in phase space:

\begin{equation}\label{vls}
\frac{\partial f_s}{\partial t} + \textbf{v}_s \cdot \nabla_r f_s  + (\frac{e_s}{m_s}) \textbf{E} \cdot \nabla_v f_s = 0,
\end{equation}
where $f_s(x, v_s, t)$ is the velocity distribution function and $\frac{e_s}{m_s}$ is the charge-to-mass ratio of the species, $s$. Meanwhile, the Poisson equation governs the electrostatic potential generated by charge distributions in a plasma: 
\begin{equation}\label{ps1}
E_x(x, t) = -\nabla \phi,
\end{equation}
\begin{equation}\label{ps2}
\Delta \phi = - \frac{\rho}{\varepsilon_0}.
\end{equation}
where $\phi(x, t)$ is the electrostatic potential and $\varepsilon_0$ is the vacuum permitivity. $\rho(x,t) = \sum_s e_s n_s$ is the charge density, where $e_s$ is the charge and  
\begin{equation}\label{ns}
n_s(x, t) = \int f_s(x, v_s, t) \ dv_s,
\end{equation}
is number density of the particle species, $s$.

Solving the Vlasov-Poisson equations using traditional numerical methods is computationally expensive due to the high dimensionality and fine resolution required. Although multi-moment fluid models offer a more efficient alternative, they fail to capture certain kinetic plasma behaviors. In contrast, machine learning-based surrogate models provide a promising compromise by significantly reducing computational cost while maintaining the accuracy necessary to model key kinetic phenomena, including nonlinear effects.

In the present work, we use DeepONets to predict the dynamical evolution of electric field energy and thus the rate of Landau damping:
\begin{equation}
\mathcal E (t) =  \int |E_x(x,t)|^2 dx   
\end{equation}

\subsection{\label{sec:setup}Data for Training and Testing}
\begin{figure*}
    \centering
    \includegraphics[width=1\textwidth]{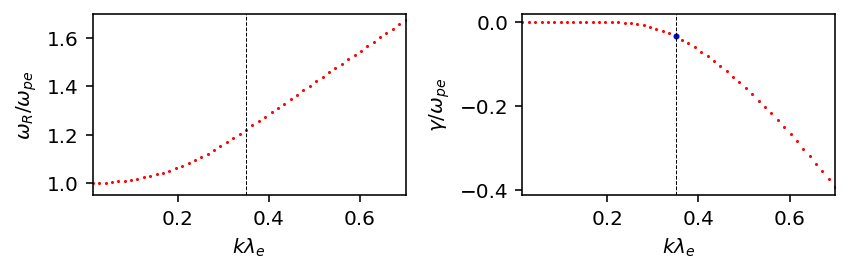}
    \caption{The dispersion relation, i.e., the real and imaginary frequencies versus wavenumber, of the least damped mode in a uniform, electrostatic plasma with immobile ion background and temperature $T = 1$. The blue dot identifies the wavenumber $k=0.35$ used for our single-mode case. The frequencies and wavenumbers are normalized over electron plasma frequency, $\omega_{pe}$, and electron Debye length, $\lambda_e$, respectively.}
    \label{fig:choosing-k}
\end{figure*}

The reference solutions are obtained by using the open-source continuum Vlasov code \texttt{Gkeyll} \citep{juno2018discontinuous}. The simulation configuration is established with a fixed background of ions serving as a neutralizing background. Initially, perturbations are applied to the electron density to initiate the dynamical evolution of the system. The number densities of ions and electrons are described as follows:
\begin{equation}
n_i(x)\equiv n_0,
\end{equation}

\begin{equation}
n_e(x, t=0)=n_0(1+\Sigma_i A_i \cos \left(k_i x\right) ),
\end{equation}
where $n_0$ is the initial uniform number density of ions and electrons without perturbation.
$A_i$ and $k_i$ are the amplitude and wavenumber of each perturbed mode, respectively.
The dispersion relation of the least damped mode in a uniform, electrostatic plasma with immobile ion background and temperature $T = 1$ is shown in Figure~\ref{fig:choosing-k}, where the left and right panels depict the oscillatory (real) frequency, $\omega_R$, and damping rate ($\gamma$; negative growth rate), respectively, versus the wavenumber, $k$. Normalized values are used for simplicity.
As can be clearly seen, in the long wavelength limit ($k\ll\lambda_e^{-1}$), there is little damping. As the wavenumber grows past about $k\sim0.2\lambda_e^{-1}$, the damping rate increases rapidly, indicating that short-wavelength modes are damped more quickly.
For our study, in the single-mode case, we use $k = 0.35$ (and a small perturbation $A = 0.05$) as marked by the blue dot in the right panel of Figure~\ref{fig:choosing-k}. This serves as a convenient baseline case where the damping is substantial but not too aggressive.
For the five-mode case, several wavenumbers $k_i$ are chosen around this baseline value to include modes with relatively comparable damping rates to allow meaningful competition between each other. The $k_i$ values and the corresponding perturbation magnitudes, $A_i$, are described in Table. \ref{Table:modeinfo}.

\begin{table}
\centering
\caption{Wavenumber and Perturbation Amplitude of Each Mode}
\begin{tabular}{c   c    c}
\hline
$n$ & $k\lambda_{e}$    & $A_n$     \\ 
\hline
1 & 0.4  & 0.1   \\ 
2 & 0.35 & 0.05  \\ 
3 & 0.25 & 0.025 \\ 
4 & 0.5  & 0.25  \\ 
5 & 0.7  & 0.5   \\ 
\hline
\end{tabular}
\label{Table:modeinfo}
\end{table}

\section{\textbf{Results and Discussion}}
In this section, we demonstrate the capability of DeepONets for two different cases. By using DeepONets, the inputs of the branch net are temperature values that are randomly chosen in the range $[0.5, 1.5]$, and the inputs of the trunk net are equidistant points $t^j = j \Delta t \ (j=0, 1, \cdots, J)$ with a step size of $\Delta t$. The output of the network is the electric field energy. During the training phase, the model is optimized by minimizing the loss function to determine the weights. In the testing phase, the trained model is applied to unseen data to assess its accuracy and generalization capability. The network architecture and hyperparameters have been specified in Table \ref{tab}, and the optimization algorithm employed is the Adaptive Moment Estimation (Adam) method, with an exponential learning rate decay starting at $0.001$.\\

\begin{table*}
\centering
	\caption{
DeepONets Architectures and Hyperparameters Used for Training and Testing}
\label{tab}
\begin{tabular}{c ccccccc}
\hline
{Problems} & {Number of} & {Number of} &{Depth} & {Width} & {Activation } &  {Optimizer} & {Iterations}\\
{} & {Training Samples} & {Test Samples} &{} & {} & {Function} &  {} & {}\\
\hline
Single mode & $200$ & $50$ & $6$ & $200$  & $\tanh$ & Adam & $10^6$ \\
Five modes & $400$ & $100$ & $6$ & $200$  & $\tanh$ & Adam & $10^6$ \\ 
\hline
\hline\noalign{\smallskip}
\end{tabular}
\end{table*}

\subsection{Single-mode Case}
We first consider the initial condition consisting of a single mode with $k = 0.35$ and $A = 0.05$, and the network is trained on $t\in[0, 20\omega_{pe}^{-1}$] with a step size of $\Delta t = 0.002\omega_{pe}^{-1}$. 

\begin{figure*}[htbp]  
    \includegraphics[width=1.0\textwidth]{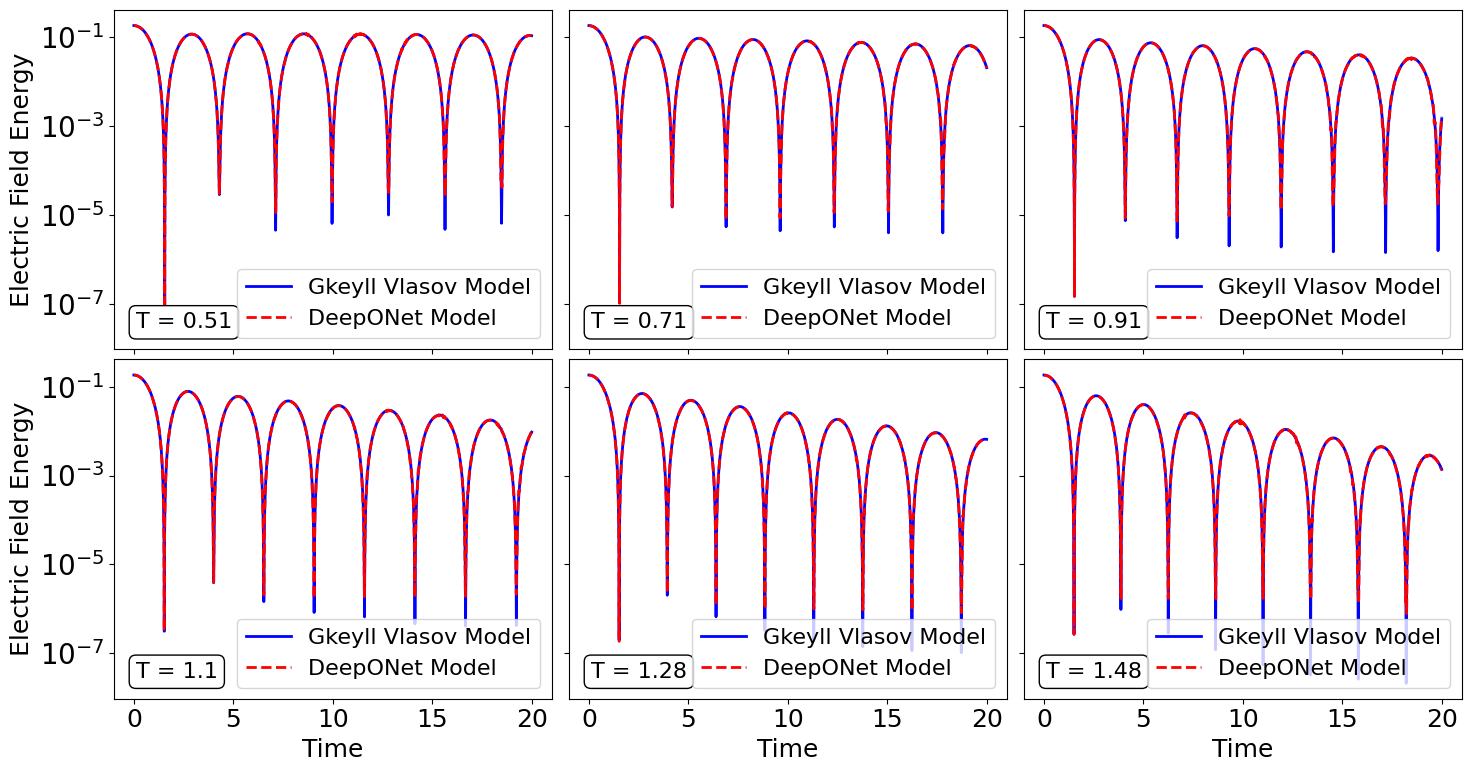}                                              
    \caption{Single-mode case: comparison of DeepONets predictions with reference solutions obtained from the Gkeyll Vlasov model for a range of temperature values ($T$). The vertical axis shows the electric field energy, $\int |E_x|^2 dx$, on a logarithmic scale, and the horizontal axis represents time in $\omega_{pe}^{-1}$.}
    \label{figsm}
    \end{figure*}

Figure \ref{figsm} compares the results obtained from DeepONets with the reference solutions generated by the Gkeyll Vlasov model for various previously unseen test samples. Figure \ref{figsm} shows that DeepONets are capable of accurately capturing the dynamical evolution of the system, as the predicted solutions match with the reference solutions.
    
The accuracy of the algorithm is assessed using the relative $L^2$ error norm, calculated as follows:
\begin{equation}\label{eqerrnorm}
\mathrm{relative~L^2~error} = \frac{\sqrt {\sum_{j=1}^J (\mathcal E (t^j) - G_{\omega}(T)(t^j) )^2} }{\sqrt{ \sum_{j=1}^J (\mathcal E (t^j))^2}}.
\end{equation}

Table \ref{tabm1} summarizes the error norm statistics for the single-mode case, including the mean, minimum, maximum, and standard deviation of the errors. The results indicate that DeepONets perform well, achieving a mean error norm of $0.0078$ for training and $0.0083$ for testing. Additionally, the small standard deviations ($0.00215$ for training and $0.00220$ for testing) demonstrate the model's consistency in both phases. These results highlight DeepONets' stability, accuracy, and uniform performance across different data samples, which indicates that DeepONets produce reliable results with minimal fluctuation in errors.

\begin{figure}[!htbp]
    \centering
        \includegraphics[scale=0.45]{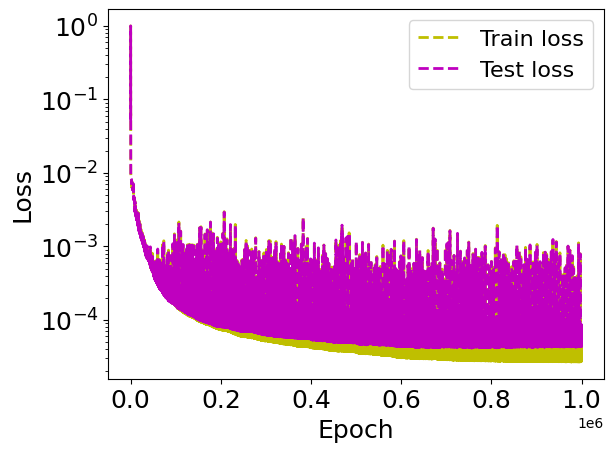}
        \caption{Single-mode case: training and test loss functions.}
    \label{sm_loss}
        \end{figure}

\begin{table}
\centering
	\caption{Statistics of Relative $L^2$ Errors for the Single-mode Case}
\label{tabm1}
\begin{tabular}{ccccc}
\hline
{} & {Mean} & {Min} & {Max} & {std. dev.}\\
\hline
Training error	& $0.0078$  & $0.0049$ & $0.0248$ & $0.00215$\\
\hline
Test error	& $0.0083$  & $0.0054$ & $0.0160$ & $0.00220$\\
\hline
\hline\noalign{\smallskip}
\end{tabular}
\end{table}  
The convergence of loss functions is depicted in Figure \ref{sm_loss}. The loss functions decrease continuously, indicating that the DeepONets surrogate model is learning effectively and converging to a solution that minimizes the error.
The loss function used is the mean square error (MSE), which is defined as follows:
\begin{equation}\label{eqloss}
\mathrm{MSE} = \frac{1}{N}\sum_{i=1}^N \sum_{j=1}^J (\mathcal E_i (t^j) - G_{\omega}(T_i)(t^j))^2.
\end{equation}
$N$ is the number of samples, $G_{\omega}(T_i)(t^j)$ is the predicted value corresponding to $T_i$ at $t^j$, and $\mathcal E_i (t^j)$ is the true value.  In the loss function, $\omega$ is used because it represents the model's trainable parameters that are adjusted to minimize the error between the predicted and the true values. This allows the model to learn and improve its predictions during the training process.
  
\subsection{Five-mode Case}
Now, let us consider the initial condition consists of five modes, with the corresponding values listed in Table \ref{Table:modeinfo}. The network is trained using $400$ samples over the interval $[0, 40\omega_{pe}^{-1}]$ with a time step of $\Delta t = 0.002\omega_{pe}^{-1}$, and the test set consists of $100$ samples.

\begin{figure*}
    \includegraphics[width=1.0\textwidth]{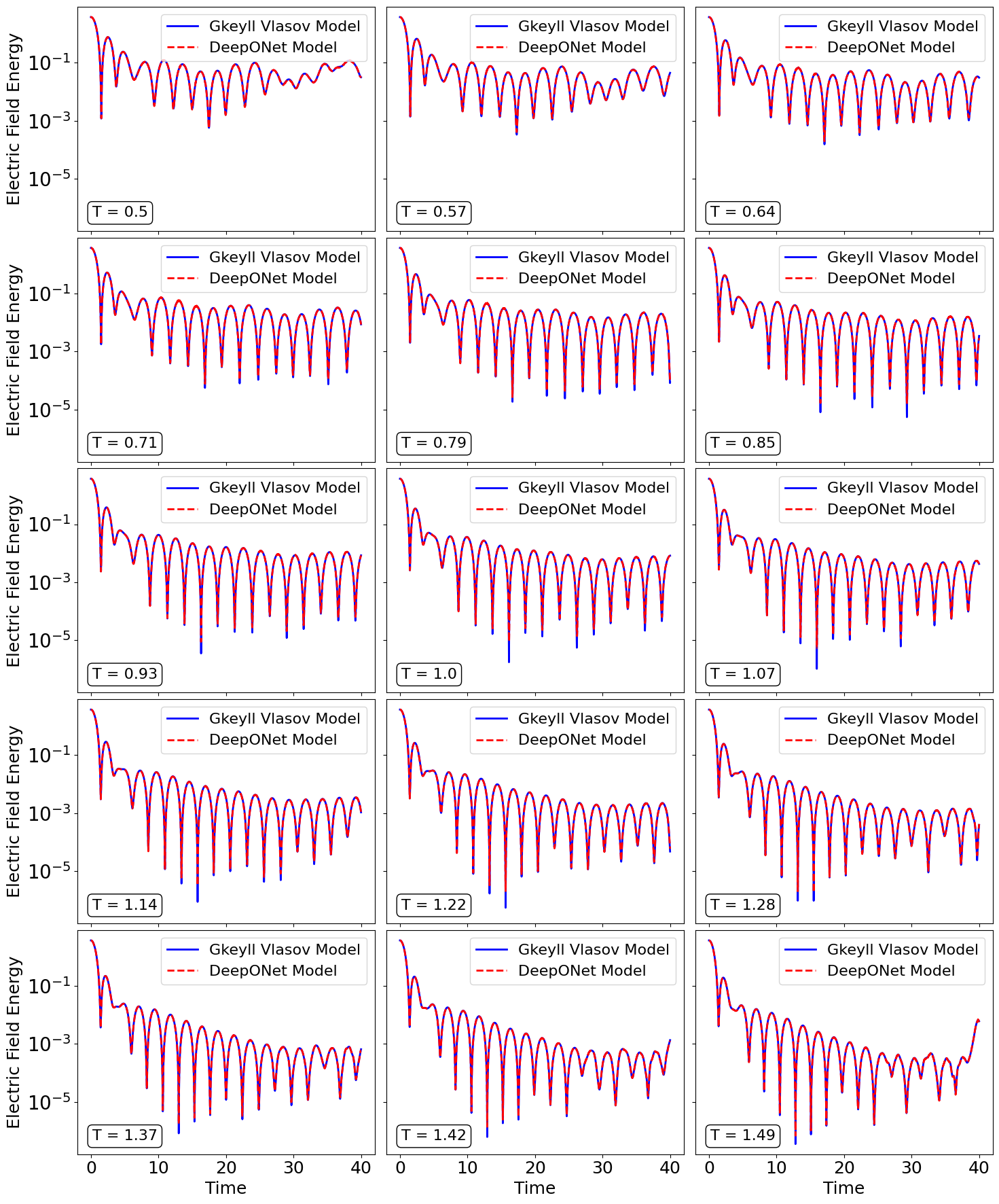}                                             
    \caption{Five-mode case: comparison between DeepONet predictions and reference Gkeyll Vlasov solutions for varying temperatures ($T$). The vertical axis shows the electric field energy, $\int |E_x|^2 dx$, on a logarithmic scale, and the horizontal axis represents time in $\omega_{pe}^{-1}$.}
\label{figfm}
\end{figure*}

Figure \ref{figfm} compares the DeepONets predictions with the corresponding reference solutions at different temperature values ($T$). As in the single-mode case, the test cases shown were not included during the training phase, highlighting the model’s generalization capability. Given the relatively large wave perturbation amplitudes, $A_n$ listed in Table \ref{Table:modeinfo} for certain modes, nonlinear Landau damping is expected to occur. In the nonlinear regime (e.g., see the top-left panel of Figure \ref{figfm}), large perturbations drive significant energy transfer from the wave to resonant particles moving slightly below the wave phase speed, accelerating them beyond it. These faster particles subsequently transfer energy back to the wave, leading to an oscillatory exchange of energy. Nonlinear Landau damping is characteristically accompanied by the formation of electron phase-space holes (see, e.g., Figure 4 in \citealp{Huang2025}). The comparison again demonstrates that the DeepONets surrogate model accurately captures the complex plasma dynamics, with its predictions in close agreement with the reference solutions.

\begin{table}[!htbp]
\centering
	\caption{Mean Relative $L^2$ Error Norms for the Five-mode Case with Different Numbers of Training and Test Samples}
\label{tabfm1}
\begin{tabular}{llll}
\hline
 {Number of} & {Number of} & {Training} & {Test}\\
{Training Samples} & {Test Samples} &{Error} & {Error} \\
\hline
\textit{$50$}	& \textit{$12$}	& $0.0035$  & $0.0312$\\
\hline 
\textit{$200$}  & \textit{$50$}	& $0.0061$ & $0.0093$\\
\hline
\textit{$400$}  & \textit{$100$} & $0.0047$ & $0.0049$\\
\hline
\textit{$800$}  & \textit{$200$}	& $0.0044$ & $0.0043$\\
\hline
\hline\noalign{\smallskip}
\end{tabular}
\end{table}

The mean relative $L^2$ error norms of the predicted solutions for different numbers of training samples are presented in Table \ref{tabfm1}. As the number of training samples increases, the test errors decrease, and the model's predictive accuracy on previously unseen test data improves, indicating its robustness and reliability. The observed increase in training error norms may be attributed to the model's exposure to more complex patterns and greater data variability with larger training sets.

\begin{figure*}
    \centering
        \includegraphics[scale=0.50]{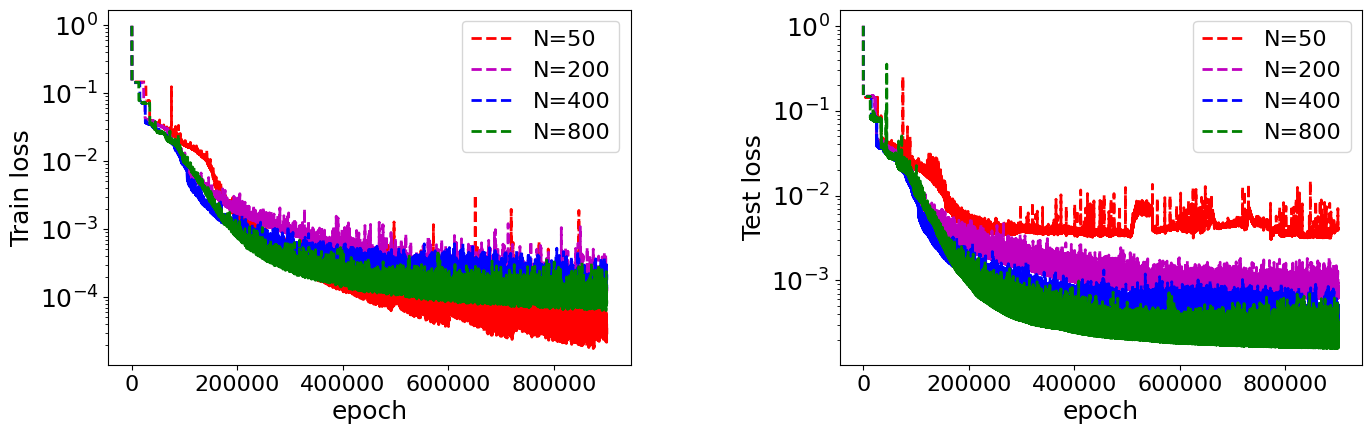}
        \caption{Five-mode case: training (left) and test (right) losses as a function of epochs for different training sample sizes (see Table \ref{tabfm1}).}
\label{lossanalfm}
\end{figure*}

The convergence of the loss functions with different numbers of training samples, depicted in Figure \ref{lossanalfm}, tends toward smaller values of $10^{-4}$. This demonstrates that the DeepONets surrogate model not only effectively learns from the training data but also improves its accuracy on test data as the size of the training dataset increases, which again highlights the model's robustness and generalization capability.

The proposed method demonstrates remarkable efficiency, with a training time of $1.93 \times 10^{-3}$s per epoch. In the best-performing run, predicting 100 test cases took only $0.00148s$ using one NVIDIA L40S GPU with 32 GB of memory, representing a significant speedup compared to the conventional numerical solver applied to the same test dataset.

\section{\textbf{Conclusion}}

The integration of machine learning techniques with plasma physics has demonstrated that data-driven approaches can effectively model and interpret complex plasma dynamics in the collisionless regime. Among these approaches, Deep Operator Networks (DeepONets) are notable for by their ability to directly approximate solutions to high-dimensional differential equations without the need for explicit discretization.

This study explores the use of DeepONets to simulate the Landau damping process - a fundamental kinetic phenomenon in space and astrophysical plasmas - highlighting their advantages over traditional numerical methods. The trained DeepONets surrogate model accurately captures the evolution of electric field energy in both single-mode and five-mode scenarios, achieving accuracy comparable to fully kinetic first-principles simulations. Its ability to generalize across varying initial conditions and perturbations underscores its robustness and adaptability. Notably, the neural network is trained on a range of initial conditions, enabling it to predict the evolution of new, unseen inputs without retraining, as long as the governing physical laws remain consistent.

By learning solution operators, DeepONets provide a computationally efficient and accurate framework for simulating complex plasma dynamics. These findings demonstrate the potential of DeepONets for broader applications in plasma physics, especially for modeling nonlinear and kinetic phenomena. Future research may extend this approach to other plasma processes and explore its performance in terms of accuracy, computational efficiency, and generalizability.

\begin{acknowledgments}
This work was partially supported by NASA grant 80NSSC23K0908, DOE grant DE-SC0024639, and the Alfred P. Sloan Research Fellowship. The authors thank Liang Wang for insightful discussions and for providing the dispersion relation calculations. We would like to acknowledge high-performance computing support from the NASA High-End Computing Program through the NASA Advanced Super-computing Division at Ames Research Center, from National Energy Research Scientific Computing Center, a DOE Office of Science user facility, and from the Derecho system (\texttt{doi:10.5065/qx9a-pg09}) provided by the NSF National Center for Atmospheric Research (NCAR), sponsored by the National Science Foundation. For distribution of the model results used in this study, please contact the corresponding authors.
\end{acknowledgments}

\begin{contribution}
C.D. supervised the project and, together with S.S., designed the research. S.S. developed the machine learning architecture based on the DeepONet model and completed the data analyses. Z.H. generated the simulation data. S.S. and C.D. wrote the manuscript. All authors reviewed and approved the final version.
\end{contribution}

\appendix

\section{Error Analysis and Convergence Behavior in the Five-Mode Case}

\begin{figure*}
    \centering
        \includegraphics[scale=0.50]{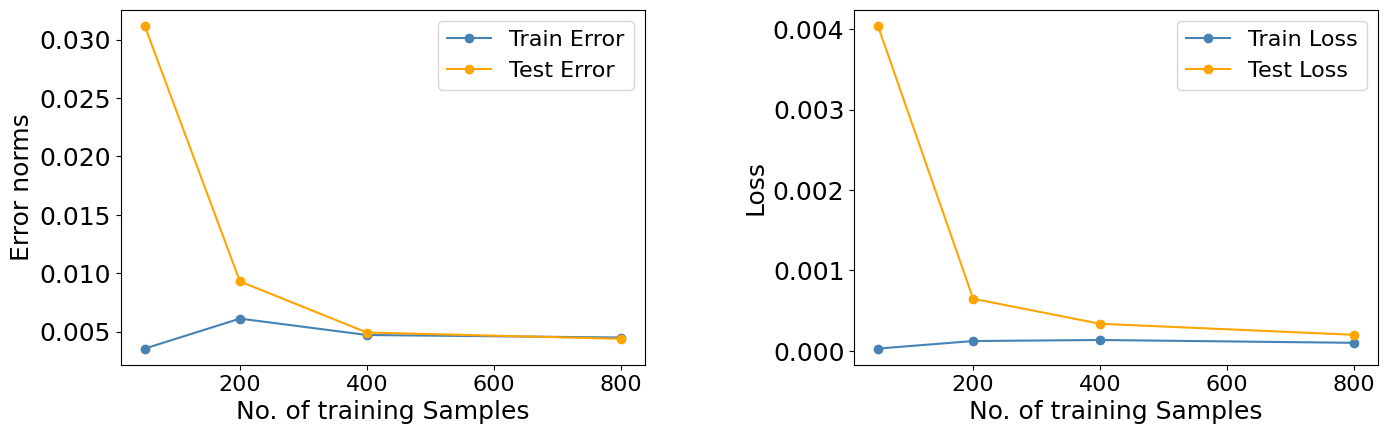}
\caption{Five-mode case: the left panel shows the error norms for training and testing as a function of the number of training samples. The right panel depicts the training and test losses vs. the number of training samples. The error norms and losses are calculated using Equations (\ref{eqerrnorm}) and (\ref{eqloss}).}
\label{errlossanalfm}
\end{figure*}

\begin{figure*}
    \centering
        \includegraphics[scale=0.47]{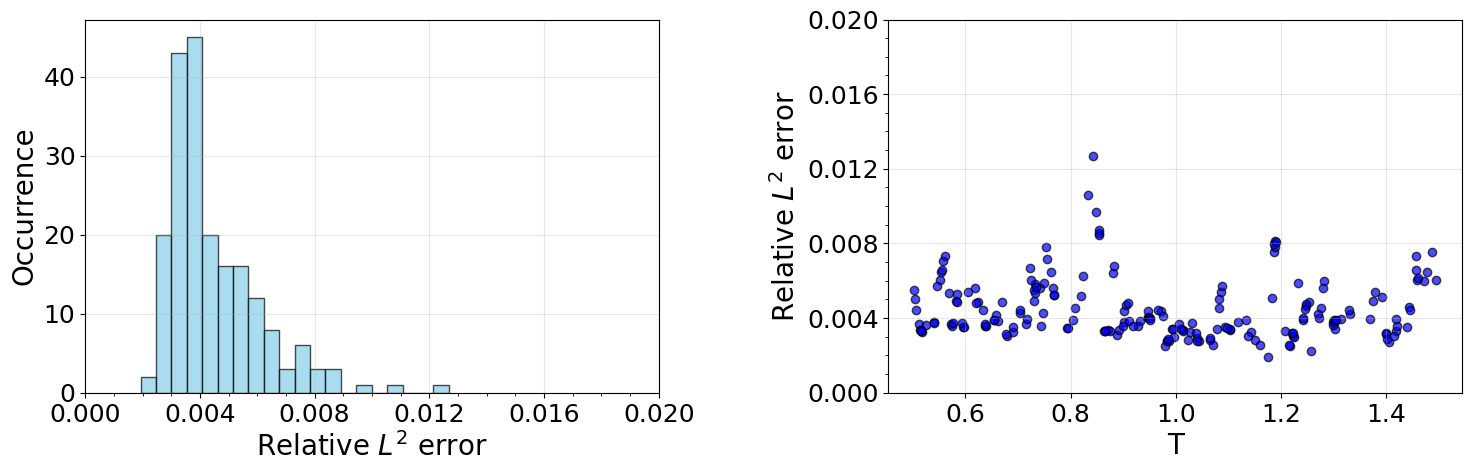}
\caption{Five-mode case: histogram (left) and scatter plot (right) of the relative $L^2$ test errors for the case with 800 training samples and 200 test samples.}
\label{hsg_fm}
\end{figure*}

In Figure \ref{errlossanalfm}, we examine how the number of training samples affects model performance in the five-mode case. Figure \ref{errlossanalfm} displays the mean relative $L^2$ error norms and loss functions for both training and test samples, plotted against the number of training samples.  After training for $400$ samples, DeepONets achieved a prediction error of less than $0.0049$. Figure \ref{hsg_fm} provides further insights into the distribution of prediction errors across all test datasets for the case with 800 training samples and 200 test samples.

\newpage

\providecommand{\noopsort}[1]{}\providecommand{\singleletter}[1]{#1}%


\begin{thebibliography}{}
\expandafter\ifx\csname natexlab\endcsname\relax\def\natexlab#1{#1}\fi
\providecommand{\url}[1]{\href{#1}{#1}}
\providecommand{\dodoi}[1]{doi:~\href{http://doi.org/#1}{\nolinkurl{#1}}}
\providecommand{\doeprint}[1]{\href{http://ascl.net/#1}{\nolinkurl{http://ascl.net/#1}}}
\providecommand{\doarXiv}[1]{\href{https://arxiv.org/abs/#1}{\nolinkurl{https://arxiv.org/abs/#1}}}

\bibitem[{E.~P. {Alves} \& F. {Fiuza}(2022){Alves} \& {Fiuza}}]{Alves2022}
{Alves}, E.~P., \& {Fiuza}, F. 2022, \bibinfo{title}{{Data-driven discovery of
  reduced plasma physics models from fully kinetic simulations},} Physical
  Review Research, 4, 033192, \dodoi{10.1103/PhysRevResearch.4.033192}

\bibitem[{M. Azizmalayeri \& M.~H. Rohban(2023)Azizmalayeri \&
  Rohban}]{Azizmalayeri2023}
Azizmalayeri, M., \& Rohban, M.~H. 2023, \bibinfo{title}{Lagrangian objective
  function leads to improved unforeseen attack generalization,} Machine
  Learning, 112, 3003, \dodoi{10.1007/s10994-023-06348-3}

\bibitem[{S. Cai {et~al.}(2021)Cai, Wang, Lu, Zaki, \& Karniadakis}]{Cai2021}
Cai, S., Wang, Z., Lu, L., Zaki, T.~A., \& Karniadakis, G. 2021,
  \bibinfo{title}{{D}eep{MMN}et: Inferring the electroconvection multiphysics
  fields based on operator approximation by neural networks,} Journal of
  Computational Physics, 436, 110296, \dodoi{10.1016/j.jcp.2021.110296}

\bibitem[{Y. {Chen} {et~al.}(2023){Chen}, {T{\'o}th}, {Zhou}, \&
  {Wang}}]{Chen2023}
{Chen}, Y., {T{\'o}th}, G., {Zhou}, H., \& {Wang}, X. 2023,
  \bibinfo{title}{{FLEKS: A flexible particle-in-cell code for multi-scale
  plasma simulations},} Computer Physics Communications, 287, 108714,
  \dodoi{10.1016/j.cpc.2023.108714}

\bibitem[{Y. Chen {et~al.}(2017)Chen, Tóth, Cassak, Jia, Gombosi, Slavin,
  Markidis, Peng, Jordanova, \& Henderson}]{Chen2017}
Chen, Y., Tóth, G., Cassak, P., {et~al.} 2017, \bibinfo{title}{Global
  Three-Dimensional Simulation of Earth's Dayside Reconnection Using a Two-Way
  Coupled Magnetohydrodynamics With Embedded Particle-in-Cell Model: Initial
  Results,} Journal of Geophysical Research: Space Physics, 122, 10,318,
  \dodoi{{10.1002/2017JA024186}}

\bibitem[{W. Cheng {et~al.}(2023)Cheng, Fu, Wang, Dong, Jin, Jiang, Ma, Qin, \&
  Liu}]{Cheng2023}
Cheng, W., Fu, H., Wang, L., {et~al.} 2023, \bibinfo{title}{Data-driven,
  multi-moment fluid modeling of Landau damping,} Computer Physics
  Communications, 282, 108538,
  \dodoi{https://doi.org/10.1016/j.cpc.2022.108538}

\bibitem[{L.~K.~S. {Daldorff} {et~al.}(2014){Daldorff}, {T{\'o}th}, {Gombosi},
  {Lapenta}, {Amaya}, {Markidis}, \& {Brackbill}}]{Daldorff2014}
{Daldorff}, L. K.~S., {T{\'o}th}, G., {Gombosi}, T.~I., {et~al.} 2014,
  \bibinfo{title}{{Two-way coupling of a global Hall magnetohydrodynamics model
  with a local implicit particle-in-cell model},} Journal of Computational
  Physics, 268, 236, \dodoi{10.1016/j.jcp.2014.03.009}

\bibitem[{J. {Donaghy} \& K. {Germaschewski}(2023){Donaghy} \&
  {Germaschewski}}]{Donaghy2023}
{Donaghy}, J., \& {Germaschewski}, K. 2023, \bibinfo{title}{{In search of a
  data-driven symbolic multi-fluid ten-moment model closure},} Journal of
  Plasma Physics, 89, 895890105, \dodoi{10.1017/S0022377823000119}

\bibitem[{C. {Dong} {et~al.}(2019){Dong}, {Wang}, {Hakim}, {Bhattacharjee},
  {Slavin}, {DiBraccio}, \& {Germaschewski}}]{Dong2019}
{Dong}, C., {Wang}, L., {Hakim}, A., {et~al.} 2019, \bibinfo{title}{{Global
  Ten-Moment Multifluid Simulations of the Solar Wind Interaction with Mercury:
  From the Planetary Conducting Core to the Dynamic Magnetosphere},}
  Geophysical Research Letters, 46, 11,584, \dodoi{10.1029/2019GL083180}

\bibitem[{Z. Hajimohammadi {et~al.}(2023)Hajimohammadi, Shekarpaz, \&
  Parand}]{Hajimohammadi2023}
Hajimohammadi, Z., Shekarpaz, S., \& Parand, K. 2023, \bibinfo{title}{The novel
  learning solutions to nonlinear differential models on a semi-infinite
  domain,} Engineering with Computers, 39, 2169–2186,
  \dodoi{10.1007/s00366-022-01603-y}

\bibitem[{G.~W. Hammett \& F.~W. Perkins(1990)Hammett \&
  Perkins}]{Hammett1990fluid}
Hammett, G.~W., \& Perkins, F.~W. 1990, \bibinfo{title}{Fluid moment models for
  Landau damping with application to the ion-temperature-gradient instability,}
  Physical Review Letters, 64, 3019, \dodoi{10.1103/PhysRevLett.64.3019}

\bibitem[{K. Hornik {et~al.}(1989)Hornik, Stinchcombe, \& White}]{Hornik1989}
Hornik, K., Stinchcombe, M., \& White, H. 1989, \bibinfo{title}{Multilayer
  feedforward networks are universal approximators,} Neural Networks, 2 (5),
  359, \dodoi{10.1016/0893-6080(89)90020-8}

\bibitem[{Z. {Huang} {et~al.}(2025){Huang}, {Dong}, \& {Wang}}]{Huang2025}
{Huang}, Z., {Dong}, C., \& {Wang}, L. 2025, \bibinfo{title}{{Machine-learning
  heat flux closure for multi-moment fluid modeling of nonlinear Landau
  damping},} Proceedings of the National Academy of Science, 122, e2419073122,
  \dodoi{10.1073/pnas.2419073122}

\bibitem[{E.~R. Ingelsten {et~al.}(2025)Ingelsten, McGrae-Menge, Alves, \&
  Pusztai}]{Ingelsten2025}
Ingelsten, E.~R., McGrae-Menge, M.~C., Alves, E.~P., \& Pusztai, I. 2025,
  \bibinfo{title}{Data-driven discovery of a heat flux closure for
  electrostatic plasma phenomena,} Journal of Plasma Physics, 91, E64,
  \dodoi{10.1017/S0022377825000285}

\bibitem[{S. {Jarmak} {et~al.}(2020){Jarmak}, {Leonard}, {Akins}, {Dahl},
  {Cremons}, {Cofield}, {Curtis}, {Dong}, {Dunham}, {Journaux}, {Murakami},
  {Ng}, {Piquette}, {Girija}, {Rink}, {Schurmeier}, {Stein}, {Tallarida},
  {Telus}, {Lowes}, {Budney}, \& {Mitchell}}]{Jarmak2020}
{Jarmak}, S., {Leonard}, E., {Akins}, A., {et~al.} 2020,
  \bibinfo{title}{{QUEST: A New Frontiers Uranus orbiter mission concept
  study},} Acta Astronautica, 170, 6, \dodoi{10.1016/j.actaastro.2020.01.030}

\bibitem[{A.~S. {Joglekar} \& A.~G.~R. {Thomas}(2023){Joglekar} \&
  {Thomas}}]{Joglekar2023}
{Joglekar}, A.~S., \& {Thomas}, A. G.~R. 2023, \bibinfo{title}{{Machine
  learning of hidden variables in multiscale fluid simulation},} Machine
  Learning: Science and Technology, 4, 035049, \dodoi{10.1088/2632-2153/acf81a}

\bibitem[{J. Juno {et~al.}(2018)Juno, Hakim, TenBarge, Shi, \&
  Dorland}]{juno2018discontinuous}
Juno, J., Hakim, A., TenBarge, J., Shi, E., \& Dorland, W. 2018,
  \bibinfo{title}{Discontinuous Galerkin algorithms for fully kinetic plasmas,}
  Journal of Computational Physics, 353, 110, \dodoi{10.1016/j.jcp.2017.10.009}

\bibitem[{T. Kurth {et~al.}(2023)Kurth, Subramanian, Harrington, Pathak,
  Mardani, Hall, Miele, Kashinath, \& Anandkumar}]{Pathak2023}
Kurth, T., Subramanian, S., Harrington, P., {et~al.} 2023, in Proceedings of
  the Platform for Advanced Scientific Computing Conference, PASC '23 (New
  York, NY, USA: Association for Computing Machinery),
  \dodoi{10.1145/3592979.3593412}

\bibitem[{B. {Laperre} {et~al.}(2022){Laperre}, {Amaya}, {Jamal}, \&
  {Lapenta}}]{Laperre2022}
{Laperre}, B., {Amaya}, J., {Jamal}, S., \& {Lapenta}, G. 2022,
  \bibinfo{title}{{Identification of high order closure terms from fully
  kinetic simulations using machine learning},} Physics of Plasmas, 29, 032706,
  \dodoi{10.1063/5.0066397}

\bibitem[{D. {Li} {et~al.}(2023){Li}, {Chen}, {Dong}, {Wang}, \&
  {Toth}}]{Li2023}
{Li}, D., {Chen}, Y., {Dong}, C., {Wang}, L., \& {Toth}, G. 2023,
  \bibinfo{title}{{Numerical study of magnetic island coalescence using
  magnetohydrodynamics with adaptively embedded particle-in-cell model},} AIP
  Advances, 13, 015126, \dodoi{10.1063/5.0122087}

\bibitem[{Z. Li {et~al.}(2020)Li, Kovachki, Azizzadenesheli, Liu, Bhattacharya,
  Stuart, \& Anandkumar}]{Li2020fourier}
Li, Z., Kovachki, N., Azizzadenesheli, K., {et~al.} 2020,
  \bibinfo{title}{Fourier neural operator for parametric partial differential
  equations,} arXiv preprint, arXiv:2010.08895,
  \dodoi{10.48550/arXiv.2010.08895}

\bibitem[{L. Lu {et~al.}(2021)Lu, Jin, Pang, Zhang, \& Karniadakis}]{Lulu2021}
Lu, L., Jin, P., Pang, G., Zhang, Z., \& Karniadakis, G. 2021,
  \bibinfo{title}{Learning nonlinear operators via {DeepONet} based on the
  universal approximation theorem of operators,} Nature Machine Intelligence,
  3, 218, \dodoi{10.1038/s42256-021-00302-5}

\bibitem[{C. Ma {et~al.}(2020)Ma, Zhu, Xu, \& Wang}]{Ma2020machine}
Ma, C., Zhu, B., Xu, X.-Q., \& Wang, W. 2020, \bibinfo{title}{Machine learning
  surrogate models for Landau fluid closure,} Physics of Plasmas, 27,
  \dodoi{/10.1063/1.5129158}

\bibitem[{S. Mao {et~al.}(2023)Mao, Dong, Lu, Yi, Wang, \&
  Perdikaris}]{Mao2023}
Mao, S., Dong, R., Lu, L., {et~al.} 2023, \bibinfo{title}{{PPDON}et: Deep
  operator networks for fast prediction of steady-state solutions in
  disk–planet systems,} The Astrophysical Journal Letters, 950, L12,
  \dodoi{10.3847/2041-8213/acd77f}

\bibitem[{X. Meng \& G. Karniadakis(2020)Meng \& Karniadakis}]{Meng2020}
Meng, X., \& Karniadakis, G. 2020, \bibinfo{title}{A composite neural network
  that learns from multi-fidelity data: Application to function approximation
  and inverse PDE problems,} Journal of Computational Physics, 401, 109020,
  \dodoi{10.1016/j.jcp.2019.109020}

\bibitem[{G. Pang {et~al.}(2019)Pang, Lu, \& Karniadakis}]{Pang2019}
Pang, G., Lu, L., \& Karniadakis, G. 2019, \bibinfo{title}{{FPINN}s: Fractional
  physics-informed neural networks,} SIAM Journal on Scientific Computing, 41,
  A2603, \dodoi{10.1137/18M1229845}

\bibitem[{K. Parand {et~al.}(2020)Parand, Razzaghi, Sahleh, \&
  Jani}]{Parand2022}
Parand, K., Razzaghi, M., Sahleh, R., \& Jani, M. 2020, \bibinfo{title}{Least
  squares support vector regression for solving Volterra integral equations,}
  Engineering with Computers, 36, 789, \dodoi{10.1007/s00366-020-01186-6}

\bibitem[{K. Parand {et~al.}(2019)Parand, Yari, Taheri, \&
  Shekarpaz}]{Parand2019}
Parand, K., Yari, H., Taheri, R., \& Shekarpaz, S. 2019, \bibinfo{title}{A
  comparison of Newton--Raphson method with Newton--Krylov generalized minimal
  residual (GMRes) method for solving one and two dimensional nonlinear
  Fredholm integral equations,} SeMA Journal, 76, 615,
  \dodoi{10.1007/s40324-019-00196-9}

\bibitem[{Y. Qin {et~al.}(2023)Qin, Ma, Jiang, Dong, Fu, Wang, Cheng, \&
  Jin}]{Qin2023}
Qin, Y., Ma, J., Jiang, M., {et~al.} 2023, \bibinfo{title}{Data-driven modeling
  of Landau damping by physics-informed neural networks,} Physical Review
  Research, 5, 033079, \dodoi{10.1103/PhysRevResearch.5.033079}

\bibitem[{M. Raissi {et~al.}(2019)Raissi, Perdikaris, \&
  Karniadakis}]{Raissi2019}
Raissi, M., Perdikaris, P., \& Karniadakis, G.~E. 2019,
  \bibinfo{title}{Physics-informed neural networks: A deep learning framework
  for solving forward and inverse problems involving nonlinear partial
  differential equations,} Journal of Computational Physics, 378, 686,
  \dodoi{10.1016/j.jcp.2018.10.045}

\bibitem[{M. Razzaghi {et~al.}(2023)Razzaghi, Shekarpaz, \&
  Rajabi}]{Razzaghi2023}
Razzaghi, M., Shekarpaz, S., \& Rajabi, A. 2023, Solving Ordinary Differential
  Equations by LS-SVM, ed. J.~A. Rad, K.~Parand, \& S.~Chakraverty (Singapore:
  Springer Nature Singapore), 147--170, \dodoi{10.1007/978-981-19-6553-1_7}

\bibitem[{E. {Rulke} {et~al.}(2021){Rulke}, {Wang}, \& {Dong}}]{Rulke2021}
{Rulke}, E., {Wang}, L., \& {Dong}, C. 2021, in AGU Fall Meeting Abstracts,
  Vol. 2021, SM53C--08

\bibitem[{I. Sahin {et~al.}(2024)Sahin, Moya, Mollaali, Lin, \&
  Paniagua}]{Sahin2024}
Sahin, I., Moya, C., Mollaali, A., Lin, G., \& Paniagua, G. 2024,
  \bibinfo{title}{Deep operator learning-based surrogate models with
  uncertainty quantification for optimizing internal cooling channel rib
  profiles,} International Journal of Heat and Mass Transfer, 219, 124813,
  \dodoi{10.1016/j.ijheatmasstransfer.2023.124813}

\bibitem[{S. Shekarpaz \& H. Azari(2020)Shekarpaz \& Azari}]{Shekarpaz2020}
Shekarpaz, S., \& Azari, H. 2020, \bibinfo{title}{Splitting method for an
  inverse source problem in parabolic differential equations: Error analysis
  and applications,} Numerical Methods for Partial Differential Equations, 36,
  654, \dodoi{10.1002/num.22447}

\bibitem[{S. Shekarpaz {et~al.}(2022)Shekarpaz, Azizmalayeri, \&
  Rohban}]{Shekarpaz2022}
Shekarpaz, S., Azizmalayeri, M., \& Rohban, M.~H. 2022, \bibinfo{title}{PIAT:
  Physics Informed Adversarial Training for Solving Partial Differential
  Equations,} \doarXiv{2207.06647}

\bibitem[{S. Shekarpaz {et~al.}(2024)Shekarpaz, Zeng, \&
  Karniadakis}]{Shekarpaz2024}
Shekarpaz, S., Zeng, F., \& Karniadakis, G. 2024, \bibinfo{title}{Splitting
  Physics-Informed Neural Networks for Inferring the Dynamics of Integer- and
  Fractional-Order Neuron Models,} Communications in Computational Physics, 35,
  1–37, \dodoi{10.4208/cicp.oa-2023-0121}

\bibitem[{K. Shukla {et~al.}(2025)Shukla, Zou, Chan, Pandey, Wang, \&
  Karniadakis}]{Shukla2025}
Shukla, K., Zou, Z., Chan, C.~H., {et~al.} 2025, \bibinfo{title}{NeuroSEM: A
  hybrid framework for simulating multiphysics problems by coupling PINNs and
  spectral elements,} Computer Methods in Applied Mechanics and Engineering,
  433, 117498, \dodoi{https://doi.org/10.1016/j.cma.2024.117498}

\bibitem[{G. {T{\'o}th} {et~al.}(2016){T{\'o}th}, {Jia}, {Markidis}, {Peng},
  {Chen}, {Daldorff}, {Tenishev}, {Borovikov}, {Haiducek}, {Gombosi}, {Glocer},
  \& {Dorelli}}]{Toth2016}
{T{\'o}th}, G., {Jia}, X., {Markidis}, S., {et~al.} 2016,
  \bibinfo{title}{{Extended magnetohydrodynamics with embedded particle-in-cell
  simulation of Ganymede's magnetosphere},} Journal of Geophysical Research
  (Space Physics), 121, 1273, \dodoi{10.1002/2015JA021997}

\bibitem[{L. {Wang} {et~al.}(2018){Wang}, {Germaschewski}, {Hakim}, {Dong},
  {Raeder}, \& {Bhattacharjee}}]{Wang2018}
{Wang}, L., {Germaschewski}, K., {Hakim}, A., {et~al.} 2018,
  \bibinfo{title}{{Electron Physics in 3-D Two-Fluid 10-Moment Modeling of
  Ganymede's Magnetosphere},} Journal of Geophysical Research (Space Physics),
  123, 2815, \dodoi{10.1002/2017JA024761}

\bibitem[{L. Wang {et~al.}(2015)Wang, Hakim, Bhattacharjee, \&
  Germaschewski}]{Wang2015}
Wang, L., Hakim, A.~H., Bhattacharjee, A., \& Germaschewski, K. 2015,
  \bibinfo{title}{Comparison of multi-fluid moment models with Particle-in-Cell
  simulations of collisionless magnetic reconnection,} Physics of Plasmas, 22,
  012108, \dodoi{{10.1063/1.4906063}}

\bibitem[{L. {Wang} {et~al.}(2020){Wang}, {Hakim}, {Ng}, {Dong}, \&
  {Germaschewski}}]{Wang2020}
{Wang}, L., {Hakim}, A.~H., {Ng}, J., {Dong}, C., \& {Germaschewski}, K. 2020,
  \bibinfo{title}{{Exact and locally implicit source term solvers for
  multifluid-Maxwell systems},} Journal of Computational Physics, 415, 109510,
  \dodoi{10.1016/j.jcp.2020.109510}

\bibitem[{X. Wang {et~al.}(2022)Wang, Chen, \& Tóth}]{Wang2022}
Wang, X., Chen, Y., \& Tóth, G. 2022, \bibinfo{title}{Global
  Magnetohydrodynamic Magnetosphere Simulation With an Adaptively Embedded
  Particle-In-Cell Model,} Journal of Geophysical Research: Space Physics, 127,
  e2021JA030091, \dodoi{10.1029/2021JA030091}

\bibitem[{S. Wei {et~al.}(2023)Wei, Liu, Fu, Dong, \& Wang}]{Wei2023}
Wei, S., Liu, Y., Fu, H., Dong, C., \& Wang, L. 2023, in 2023 International
  Applied Computational Electromagnetics Society Symposium (ACES-China), IEEE,
  01--03, \dodoi{10.23919/ACES-China60289.2023.10249492}

\bibitem[{D. Zhang {et~al.}(2020)Zhang, Guo, \& Karniadakis}]{Zhang2020-2}
Zhang, D., Guo, L., \& Karniadakis, G. 2020, \bibinfo{title}{Learning in modal
  space: Solving time-dependent stochastic PDEs using physics-informed neural
  networks,} SIAM Journal on Scientific Computing, 42, A639,
  \dodoi{10.1137/19M1260141}

\bibitem[{H. {Zhou} {et~al.}(2020){Zhou}, {T{\'o}th}, {Jia}, \&
  {Chen}}]{Zhou2020}
{Zhou}, H., {T{\'o}th}, G., {Jia}, X., \& {Chen}, Y. 2020,
  \bibinfo{title}{{Reconnection-Driven Dynamics at Ganymede's Upstream
  Magnetosphere: 3-D Global Hall MHD and MHD-EPIC Simulations},} Journal of
  Geophysical Research (Space Physics), 125, e28162,
  \dodoi{10.1029/2020JA028162}

\end{thebibliography}
\end{document}